# Probing valley population imbalance in transition metal dichalcogenides via temperature-dependent second harmonic generation imaging

Leonidas Mouchliadis[1]*, Sotiris Psilodimitrakopoulos[1], George Miltos Maragkakis[1,2], Ioanna Demeridou[1,2], George Kourmoulakis[1,3], Andreas Lemonis[1], George Kioseoglou[1,3] and Emmanuel Stratakis[1,2]*

[1]*Institute of Electronic Structure and Laser-Foundation for Research and Technology-Hellas, Heraklion, Greece*

[2]*Physics Department, University of Crete, Heraklion, Greece*

[3]*Department of Materials Science and Technology, University of Crete, Heraklion, Greece*

**Abstract**

Degenerate minima in momentum space –valleys– provide an additional degree of freedom that can be used for information transport and storage. Notably, such minima naturally exist in the band structure of transition metal dichalcogenides (TMDs). When these atomically thin crystals interact with intense laser light, the second harmonic generated (SHG) field inherits special characteristics that reflect not only the broken inversion symmetry in real space, but also the valley anisotropy in reciprocal space. The latter is present whenever there exists a valley population imbalance (VPI) between the two valleys. In this work, it is shown that the temperature-induced changes of the SHG intensity dependence on the excitation field polarization, is a unique fingerprint of VPI in TMDs. Analysis of such changes, in particular, enables the calculation of the valley-induced to intrinsic second order susceptibilities ratio. Unlike temperature-dependent photoluminescence (PL) measurements of valley polarization and coherence, the proposed polarization-resolved SHG (PSHG) methodology is insensitive to the excitation field wavelength, an advantage that renders it ideal for monitoring VPI in large crystalline or stacked areas comprising different TMDs.

**Introduction**

Conventional optoelectronics is based on the manipulation of electronic charge with light, for information transport, storage and readout. In electronic systems with degenerate minima in their band structure –valleys– an additional degree of freedom that labels those minima, i.e., the valley index, can serve as the information carrier. Such a possibility has opened a new field of electronics, namely valleytronics, which enables the processing of additional information within the same physical space[1, 2, 3]. Specifically, the valley index can be mapped to a pseudospin that, similar to the response of a spin to magnetic fields, is affected by the Berry curvature of the bands[4, 5].

In TMDs, the valley pseudospin is coupled to the electron spin[6] to give rise to selection rules that allow the excitation of carriers in specific valleys only when light of suitable helicity is used[7, 8, 9]. More specifically, the excitation and control of carriers is achieved by circularly polarized light that populates only one of the two valleys. Interestingly, the polarization is transferred to and subsequently measured in the emitted (one- and two-photon) PL[10, 11, 12]. Such measurements are used to indirectly probe the population imbalance in different valleys.

Several other phenomena, associated with the existence of valleys and related with time reversal symmetry, have been reported in atomically thin crystals with hexagonal structure and corresponding degenerate (but inequivalent) K and K' valleys in the hexagonal Brillouin zone. For example, in the presence of external fields, this imbalance may generate valley currents that give rise to valley Hall conductivity[13, 14]. In addition, when linearly polarized light is used for excitation, superpositions of excitons in different valleys are created and one can measure valley coherence[15, 16]. Valley injection and transport may be used to create a valley filter, i.e., a means to populate a single valley and therefore induce valley polarization. Serial combination of two such filters acts as a valley valve that can be externally controlled[18]. Moreover, although intervalley scattering is suppressed in ideal crystals, there are recent reports of intervalley collective modes in the presence of unequal valley populations[19].

In all the aforementioned cases, the common underlying physical principle is the population imbalance between different valleys. This is reasonable considering that polarization and transport effects are associated with charge separation and local variations in the chemical potential, respectively. Hence, they directly reflect the crystal symmetries in both real and momentum space. In real space, the charge is locally accumulated around the atomic positions, whereas in momentum space, the carriers occupy states in the vicinity of high symmetry points within the hexagonal Brillouin zone. Interestingly, both polarization and transport effects can be examined by analyzing the second order nonlinear optical response of atomically thin crystals.

Recently, SHG spectroscopy has been highly appreciated as a powerful tool to study two-dimensional (2D) TMDs[20, 21, 22, 23, 24, 25, 26, 27]. Owing to the vanishing thickness of these thin crystals, phase matching conditions are readily satisfied and thus the second order nonlinear susceptibility, $\chi^{(2)}$, is large[28]. Under the $D_{3h}$ point group symmetry of TMD crystals, the $\chi^{(2)}$ tensor shows non-vanishing elements for odd-layer TMDs in 2H configuration and for arbitrary-layered in 3R stacking geometry[28]. These elements are not independent but are inter-related through: $\chi^{(2)}_{xxx} = -\chi^{(2)}_{xyy} = -\chi^{(2)}_{yyx} = -\chi^{(2)}_{yxy} = \chi^{(2)}_{int}$, where $x$ denotes the crystal mirror symmetry axis, i.e., the armchair direction and $\chi^{(2)}_{int}$ is the nonzero element of the intrinsic second order susceptibility tensor.

In addition, both SHG intensity and polarization have been found to reveal information about the main crystallographic axis[20, 21, 25, 26] grain boundaries[29, 30], stacking sequence, twist angle[24, 27], number of layers and crystal homogeneity[26]. Indeed, the ability to control the polarization state of the SHG signal enables the extraction of additional information from a single measurement, since the second order response is determined by a third rank susceptibility tensor[31]; therefore measurements at higher order of response enable access to a larger number of independent quantities of a system[32, 33].

Recent theoretical studies suggest that the second order optical response is also a useful tool to probe the electronic configuration of 2D crystals[32, 34, 35]. This is possible due to the symmetry characterizing the hexagonal Brillouin zone in momentum space. Similarly to the alternating atoms at the corners of the hexagon in real space, characterized by the $D_{3h}$ symmetry of the trigonal prismatic structure with Bernal stacking (Fig. 1a), in momentum space the alternating K and K' points also result in $D_{3h}$ symmetry, reflecting the trigonal warping of electrons in the vicinity of high symmetry points (Fig.1b)[36]. Besides, the conduction and valence band states at the corners of the hexagonal Brillouin zone are formed by hybridization of the transition metal *d*-orbitals with the chalcogen *p*-orbitals and therefore are strongly localized in the metal atom plane[37, 38, 39]. A direct consequence of this effect is the possibility for such crystals to produce valley-induced SHG, additionally to the intrinsic second order response. As a result, in the presence of population imbalance between the two valleys, additional elements in the second order nonlinear optical susceptibility tensor become nonzero[32, 34, 35]. Therefore, we have: $\chi^{(2)}_{yyy} = -\chi^{(2)}_{yxx} = -\chi^{(2)}_{xxy} = -\chi^{(2)}_{xyx} = \chi^{(2)}_{vpi}$, where $\chi^{(2)}_{vpi}$ is the nonzero element of the VPI second order susceptibility tensor.

VPI can be either induced[10] or emerge spontaneously[40] and depending on the way it is created, it reflects different aspects of the electronic system. For instance, when circularly polarized light is used for excitation, only one of the two valleys is populated and the spin-valley polarization is transferred to the detected one- or two-photon PL[15]. In contrast, for linearly polarized light, a superposition of excitons in the K and K´ valleys is created and one can measure valley coherence[16]. Even at equilibrium, i.e., valley population balance, there is a small, but finite probability that an electron will be transferred to the adjacent valley, while at the same time flipping its spin, thus perturbing the balance. In this last case, access to the degree of VPI would reveal information about the intrinsic valley relaxation time[40].

In this work, we take advantage of the temperature dependence of PSHG to account for the VPI in atomically thin TMDs. Notably, since the VPI defines the difference between the valley populations, $\delta n = n_K - n_{K'}$, it also reflects the chemical potential difference, $\delta \mu$, between the two valleys (Fig. 1c). Hence, in the presence of imbalance, the additional valley-induced contribution to the SHG that is intrinsically generated by the TMD crystal, can be estimated as

$$I_{SHG}^{VPI} \sim \delta n^2 \sim \delta \mu^2$$

with the corresponding contribution to the second order nonlinear susceptibility being proportional to the chemical potential difference, i.e., $\chi_{vpi}^{(2)} \sim \delta \mu$[34, 35, 40]. Accordingly, local variations in the chemical potential affect the SHG induced by the VPI and thus can be probed by nonlinear optical experiments[41]. More importantly, and in contrast to the intrinsic nonlinear optical response of 2D TMDs, the valley-induced SHG is sensitive to temperature variations. Based on this, we here vary the temperature of a 2D TMD crystal and the corresponding changes in the SHG intensity are used to probe the inter-valley chemical potential difference and therefore the VPI. Unlike temperature-dependent PL measurements of valley polarization and coherence, the proposed SHG methodology is insensitive to the excitation field wavelength, an advantage that renders it suitable for monitoring VPI in TMDs.

We consider an electromagnetic field that is normally incident to a 2D TMD sample with polarization parallel to the sample plane, at an angle $\varphi$ (Fig. 2a); the crystal armchair direction is oriented at an angle $\theta$. Here we implement the experiment proposed by Hipolito and Pereira[32], in which a quarter-wave plate is placed before the sample with its fast axis at an angle $\alpha$. Using a half-wave plate we control the orientation $\varphi$ of the fundamental linear polarization, while a linear polarizer placed before the detector at an angle $\zeta$, selects suitable SHG components. All angles are measured with respect to the laboratory x-axis.

PSHG imaging was performed in 78K–300K temperature range using a fs laser-scanning microscope coupled with a liquid nitrogen cryogenic system (Fig. 2b; see also Methods). The infrared laser beam is guided into an inverted microscope, while its polarization is controlled by properly rotating half- and quarter-wave plates. A pair of galvanometric mirrors enables raster-scanning of the sample, which is placed inside a continuous flow cryostat. The SHG signal from the sample passes through a rotating linear polarizer and is collected in reflection geometry. Different rotation speeds of the optical elements allows control over the angles $\varphi$, $\alpha$ and $\zeta$. The armchair direction $\theta$ can be determined with the same experimental setup[26].

For the PL measurements, a micro-PL setup was used to collect PL in a backscattering geometry (see also Methods). Emitted light was dispersed by a single monochromator equipped with a multichannel CCD detector. Following the excitation, the emitted PL spectra were analyzed as $\sigma^+$ and $\sigma^-$ using a combination of quarter-wave plate and linear polarizer placed in front of the spectrometer entrance slit. A cryogenic system was coupled with the optical setups to perform temperature-dependent second harmonic and spin-valley polarization measurements in a range of temperatures from 78K up to 300K.

Application of nonlinear optics for a crystal with D$_{3h}$ symmetry yields the SHG field emerging from the crystal, as[42, 43]

$$\begin{pmatrix} P_x^{2\omega} \\ P_y^{2\omega} \end{pmatrix} = \varepsilon_0 \begin{pmatrix} \chi_{int}^{(2)}(E_x^2 - E_y^2) - 2\chi_{vpi}^{(2)} E_x E_y \\ -\chi_{vpi}^{(2)}(E_x^2 - E_y^2) - 2\chi_{int}^{(2)} E_x E_y \end{pmatrix} \qquad (1)$$

Here $\chi_{int}^{(2)}$ and $\chi_{vpi}^{(2)}$ correspond to the intrinsic and induced due to VPI contributions to the second order response, respectively. This means that the SHG intensity reaching the detector depends on four angles, namely $\varphi$, $\theta$, $\alpha$, $\zeta$, corresponding to the effects of excitation linear polarization, crystal orientation, quarter wave plate and linear polarizer, respectively (Fig. 2). In this case, the detected SHG intensity is given by:

$$I_{2\omega} = [\cos(2\alpha + \zeta - 3\theta) - \kappa \sin(2\alpha + \zeta - 3\theta)]^2 + [\kappa \cos(2\alpha + \zeta - 3\theta) + \sin(2\alpha + \zeta - 3\theta)]^2 \sin^2[2(\alpha - \varphi)] \qquad (2)$$

where κ denotes the absolute value of the valley-induced to intrinsic susceptibility ratio.

$$\kappa = \frac{|\chi_{vpi}^{(2)}|}{|\chi_{int}^{(2)}|} \qquad (3)$$

The ratio κ can be extracted upon fitting of the experimentally measured SHG intensity with equation (2) and reflects the degree of VPI. In our experiment the optical elements are controlled using motorized stages synchronized to rotate in phase and therefore $\varphi = \alpha = \zeta$, giving rise to a six-petal pattern for the PSHG intensity. With this experimental configuration, the normalized SHG intensity recorded at the detector reads

$$I_{2\omega}= \left[ \cos(3(\theta - \varphi)) - \kappa \sin(3(\theta - \varphi)) \right]^2 \qquad (4)$$

**Results and Discussion**

WS$_2$ samples were prepared with mechanical exfoliation and characterized using Raman mapping (see Methods). Fig. 3(a) shows an optical image of the sample where the monolayer (1L) region is indicated. In order to quantify the VPI, we fit the experimentally measured SHG intensity at each temperature with equation (4) to extract the dimensionless parameter κ. For 300 K, in particular, we assume that the valley-induced SHG is negligible (κ=0) and use the same equation to determine the armchair orientation $\theta$. As shown in Figures 3(b) and (c), the SHG intensity, as well as the VPI mapping of WS$_2$ at 78K appear to be relatively uniform across the sample area yielding a value of κ=0.1. However, several points of the flake boundaries correspond to increased or decreased κ values, most probably originating from local field effects that affect the electron distribution in the valleys, hence the VPI.

In Figures 3(d)-(i) we present polar plots of the SHG emerging from the same monolayer region, as a function of temperature, ranging from 78 K to 300 K. The effect of low temperature is to preserve VPI by hindering the relaxation processes due to scarcity of phonons[44]; this effect is readily imprinted onto the PSHG patterns. As a consequence, as the temperature rises the PSHG intensity becomes progressively lower (see also Fig. 4), indicating the suppression of VPI.

The experimental data (red spheres) –obtained from representative monolayer regions of interest– are fitted with equation (4) (blue line) and the theoretically predicted features of the polar patterns[32] are indeed identified: (a) increase in intensity at low tempera tures where we expect larger values of population imbalance, (b) rotation of the low-temperature polar diagram -with respect to the ambient temperature one- as it can no longer be exclusively associated with the armchair direction. These features are summarized in Fig. 4a, where the fitted SHG polar diagrams for the minimum and maximum temperature are compared.

The monotonic decrease of the SHG with temperature (Fig.4b) is reflected in the temperature dependence of κ-ratio (Fig.4c). For comparison, we present in Fig. 4d the degree of spin-valley polarization as function of temperature for the neutral and charged excitons, obtained from PL measurements performed in the same sample (see also Supplementary Information). It is evident that the degrees of VPI and spin-valley polarization both exhibit a monotonic reduction with temperature. We attribute both effects to the balancing of the populations in different valleys at elevated temperatures.

In order to further validate our method, we examine another flake of exfoliated $WS_2$, comprising monolayer, few-layer and bulk regions (Fig. 5a). The SHG intensity from the monolayer region at 78K is higher than the multilayer one, as it is clearly shown in Fig. 5b. In contrast, the VPI, depicted in Fig. 5c in terms of the parameter κ, is higher in the few-layer region. This is actually expected, since there is an additional contribution to the valley-induced SHG from the extra layers. PL measurements also demonstrate higher degree of valley polarization in the few-layer region compared to the monolayer one (Supplementary Information), further supporting the increased value of κ due to the additional layers[45, 46]. Finally, we present polar diagrams of PSHG intensity at different temperatures for monolayer (Figs. 5(d)-(f)) and few-layer regions (Figs. 5(g)-(i)). The experimental data (red spheres) –obtained from representative regions of interest– are fitted with Eq. (4) (blue line) and the fingerprints of VPI are again identified: Similar to the single layer case presented previously, as the temperature rises from 78K to 200 K, the maximum SHG intensity decreases and the polar diagram is rotated.

In contrast to PL measurements used to measure valley polarization and coherence[10, 11, 12, 15, 16, 17], VPI imaging by means of PSHG is insensitive to the excitation wavelength due to the coherent character of the underlying process. Whereas for both one- and two-photon PL imaging, access to the electronic states has to be achieved by tuning the excitation wavelength in the vicinity of the (real) excitonic resonances, such a limitation does not apply for PSHG as the scattering of two photons occurs in virtual states and no absorption is required. This advantage renders the PSHG method universally applicable for imaging VPI for the various atomically thin crystals. For instance, within the same field of view one can image the VPI of adjacent crystalline areas consisting of different 2D TMD flakes as demonstrated in Fig. 6, showing the $WS_2$ flake presented above together with a $WSe_2$ monolayer in the same field of view (Fig. 6a). Comparing the SHG from the $WS_2$ and $WSe_2$ monolayers we observe that the latter exhibits increased intensity (Fig. 6b), reflecting the larger intrinsic nonlinear susceptibility characterizing the selenides. On the contrary, the VPI imaging (Fig. 6c) and histogram (Fig. 6d) in terms of the parameter κ, reveals that $WS_2$ is found to have slightly higher κ values. Notably, apart from a quantification of the VPI, the method presented here provides an optical way to determine the monolayer nature of a TMD crystal and discriminate regions consisting of different TMDs.

More importantly, the demonstrated method is suitable for vertically stacked TMDs constituting a heterostructure, a geometry that, for PL imaging, would require two different excitation wavelengths. Such experiments, although being beyond the scope of the present work, can offer a fertile ground for future studies.

In conclusion, we have presented an all optical method based on PSHG that enables the quantification and imaging of VPI in 2D TMDs. The demonstrated approach relies in analyzing the temperature-dependent changes in the PSHG intensity pattern as a function of the incident polarization angle. Fitting of the experimental data with a nonlinear optics model, reveals the susceptibility ratio κ between the intrinsic and valley-induced response of the TMD crystal. In contrast to PL measurements of valley polarization, the PSHG-based method is insensitive to the excitation wavelength and therefore suitable for large crystalline areas containing various 2D TMDs either adjacent or in stacked geometry. It is also useful for multilayer crystals, which show no PL signal.

Considering that VPI is the fundamental physical principle behind a plethora of phenomena, including valley polarization, valley coherence, valley Hall conductivity and valley filtering, its quantification is of great importance for advances in the field of valleytronics. We envisage the work presented herein as a significant step towards this effort.

**Methods**

**Sample preparation.**

Polydimethylsiloxane films (PDMS) were fabricated from 10:1 mixing ratio of SYLGARD 182 Silicone Elastomer Kit with heat cure at 80° C for 2 hours. $WS_2$ and $WSe_2$ bulk crystals, purchased from HQ Graphene, were mechanically exfoliated on the PDMS films placed on microscope glass slides[47, 48]. Monolayers of these crystals were realized under an optical microscope and characterized with Raman spectroscopy. Then, the glass slides with PDMS films containing the monolayers were mounted on a XYZ micromechanical stage, placed under a coaxially illuminated microscope. The $WS_2$ and $WSe_2$ monolayers were transferred on $Si/SiO_2$ (285 nm) substrate using viscoelastic stamping[49].

**Optical spectroscopy measurements.**

We used a micro-PL setup with a 50x objective and appropriate filters and incorporated a liquid nitrogen-cooled cryostat to collect PL in a backscattering geometry. Emitted light was dispersed by a single monochromator equipped with a multichannel CCD detector. For optical spectroscopy measurements we used a combined μ-PL setup. A Mitutoyo 50x lens (NA: 0.42, f = 200 mm) was used to focus the excitation beam onto the sample, down to ~1 μm size. The samples were held inside a continuous flow cryostat (ST500, Jannis, USA); their exact position was controlled by a XYZ mechanical translation stage (PT3, Thorlabs, USA) and the excitation procedure was continuously monitored via a CCD optical setup. Following the excitation, the emitted PL signal passes through a long pass filter to eliminate the emission of the laser. The PL spectra were analyzed as $\sigma^+$ and $\sigma^-$ using a combination of a quarter-wave plate (liquid crystal) and a linear polarizer placed in front of the spectrometer entrance slit.

**Temperature-dependent measurements.**

A cryogenic system has been coupled with the optical setup to perform temperature-dependent second harmonic and spin-valley polarization measurements in a range of temperatures from 78K up to 300K. The cryogenic system consists of the liquid nitrogen-cooled cryostat (ST500, Janis, USA), the transfer line (Standard Flexible Transfer Line, Janis, USA), a liquid nitrogen 20lt storage dewar (Janis, USA), a temperature controller, a mechanical pump and a turbo pump.

**Raman characterization.**

Raman characterization was employed to verify the monolayer character of the samples. A Nicolet Almega XR μRaman analysis system (Thermo Scientific Instruments, Waltham MA USA) was used. The excitation wavelength was 473 nm and characterization was performed under ambient conditions. A set of data was collected from several regions across the $WS_2$ and $WSe_2$ monolayer areas (Supplementary Information, Figs. S3a, S4a). Exposure time, laser intensity and number of acquisitions were kept constant for all experiments. There is a variation in the intensity of different areas in the monolayer (Supplementary Information, Figs. S3b, S4b). For $WS_2$, analysis of the peak positions of the two most prominent Raman vibrational modes $A_1´$ (out of plane) and $E´$ (in plane), is an indicator of the monolayer character of the sample. In all cases, the difference was ranging from 59 $cm^{-1}$ to 61 $cm^{-1}$, which confirms the existence of the monolayer in all studied areas (Supplementary Information, Fig. S3c). Similarly for $WSe_2$, $E´$ (in plane) vibrational mode is positioned around 249 $cm^{-1}$ and $A_1$ (out of plane) vibrational mode is at 260 $cm^{-1}$ (Supplementary Information, Fig. S4c). Raman mapping revealed a difference of 11 $cm^{-1}$, which also indicates monolayer thickness for $WSe_2$. Finally, Raman characterization of the $WS_2$ sample comprising monolayer, few-layer and bulk areas (Supplementary Information, Fig. S5a) was performed at liquid nitrogen conditions, with an excitation energy of 543nm. The corresponding Raman spectra for the three areas, (Supplementary Information, Figs. S5d-f), suggest that they comprise one, two to three and above six layers, respectively.

**Polarization-resolved second harmonic generation imaging.**

PSHG imaging was performed using a fs laser raster-scanning microscope coupled with a liquid nitrogen cryogenic system (ST500, Janis, USA) (Fig. 2b). The laser beam from a diode-pumped Yb:KGW fs oscillator (1030 nm, 70–90 fs, 76 MHz, Pharos-SP, Light Conversion, Lithuania) is guided into an inverted microscope (modified Axio Observer Z1, Carl Zeiss, Germany). In order to control the polarization of the fundamental field, we use two motorized rotation stages of high accuracy (0.1º) (M-060.DG, Physik Instrumente, Karlsruhe, Germany), holding a half-wave plate (QWPO-1030-10-2, CVI Laser, USA) and a quarter-wave plate (QWPO-1030-10-4, CVI Laser, USA). A pair of galvanometric mirrors (6215H, Cambridge Technology, Bedford, MA, USA) directs the beam into the microscope, enabling raster-scanning of the stationary sample placed inside a continuous flow cryostat (ST-500, Janis, USA). Then, a pair of lenses forming a telescope suitably expands the beam to fill the back aperture of the microscope objective lens (50x, 0.55 NA, M Plan Apo, Mitutoyo, Japan). Given that the cryostat is not permeable, we collect the SHG in the backward (epi-) direction, with the same objective used for excitation and a short-pass dichroic mirror (DMSP805R, Thorlabs, USA) at the microscope turret box. The SHG signal is then filtered by suitable short-pass (FF01-680/SP, Semrock, Rochester, NY, USA) and narrow bandpass filters (FF01-514/3, Semrock, Rochester, NY, USA), to separate it from fundamental and absorption-induced radiation. Finally, SHG passes through a linear polarizer (LPVIS100-MP, ThorLabs, USA) placed on a third motorized rotation stage in front of the detector, which is based on a PMT module (H9305-04, Hamamatsu, Hamamatsu City, Japan). Coordination of PMT recordings with the galvanometric mirrors for the image formation, as well as the movements of all motors, are carried out using LabView (National Instruments, USA).

**Data availability statement**

The data that support the findings of this study are available from the corresponding authors upon reasonable request.

**Acknowledgements**

We would like to acknowledge financial support by the Hellenic Foundation for Research and Innovation (HFRI) under the "First Call for HFRI Research Projects to support Faculty members and Researchers and the procurement of high-cost research equipment grant" (Project No. HFRI-FM17-3034). This research has been co-financed by the European Union and Greek national funds through the Operational Program Competitiveness, Entrepreneurship and Innovation, under the call European R & T Cooperation-Grant Act of Hellenic Institutions that have successfully participated in Joint Calls for Proposals of European Networks ERA NETS (National project code:

GRAPH-EYE T8EPA2-00009 and European code: 26632, FLAGERA). Support by the European Research Infrastructure NFFA-Europe, funded by the EU's H2020 framework program for research and innovation under grant agreement no. 654360, is also gratefully acknowledged.

**Author contributions**

L.M., S.P., E.S. and G.Ki. planned the project; S.P., L.M. and G.M.M. designed the experiment; S.P., G.M.M. and I.D. conducted the optical experiments; L.M., S.P. and G.M.M. conducted the data analysis; A.L. provided technical support; L.M. elaborated the theoretical model; G.Ko. prepared the samples; E.S. and G.Ki. guided the research. All authors contributed to the discussion and preparation of the manuscript.

**Additional Information**

Supplementary information is available for this paper.

**Correspondence and requests for materials should be addressed to** Leonidas Mouchliadis or Emmanuel Stratakis

**Competing interests**

The authors declare no competing interests.

**Figure legends**

**Fig. 1 Crystal structure, Brillouin zone and dispersion of monolayer transition metal dichalcogenides.**

**a** Top view of monolayer TMD honeycomb lattice with transition metal atoms in red and chalcogen atoms in blue. The lattice vectors $a_1$ and $a_2$, and first-neighbor vectors $\delta_1$, $\delta_2$ and $\delta_3$ are also shown. The region bounded by dashed lines corresponds to the primitive cell. **b** First Billouin zone (shaded) of monolayer TMD showing the high symmetry points $\Gamma$, K and K'. The reciprocal lattice vectors $b_1$ and $b_2$ are also depicted. **c** Schematic of monolayer TMD dispersion. The paraboloids represent valence and conduction bands at K (purple) and K' (cyan) points. The white dashed lines indicate the chemical potential at different valleys for the case of population imbalance.

**Fig. 2 Experimental configuration and setup.**

**a** Cartoon of the experimental configuration. The definition of the angles $\varphi$, $\alpha$, $\theta$ and $\zeta$ is shown, in terms of the laboratory coordinate system (x, y black axes) and the fast axes (dashed lines) of the optical elements. **b** Sketch of the experimental setup. An infrared fs laser excites a 2D TMD held inside the cryostat. The SHG signal emerging from the sample is directed to the detector in reflection geometry. HWP: half-waveplate; QWP: quarter-waveplate; GM: galvanometric mirrors; L1, L2: lenses; D: dichroic mirror; O: objective lens; SP: sample plane; M: mirror; SPF: short-pass filter; BPF: band-pass filter; LP: linear polarizer; PMT: photomultiplier tube.

**Fig. 3 Temperature-dependent imaging of valley population imbalance in monolayer WS$_2$**

**a** Optical image of the WS$_2$ sample  **b** SHG intensity mapping of monolayer WS$_2$ at 78 K **c** Imaging of VPI in terms of parameter κ, at 78 K **d-i** PSHG intensity polar diagrams as function of the incident field polarization angle, φ, for temperatures 78, 100, 150, 200, 250 and 300 K, respectively. The red spheres correspond to experimental data and the blue line is the fitting with equation (4).

**Fig. 4 Temperature dependence of second harmonic generation intensity, valley population imbalance and spin-valley polarization.**

**a** Comparison between the SHG polar diagrams fitted to the experimental data, for the highest (red) and lowest (blue) considered temperatures. With decreasing temperature, the SHG intensity increases due to the valley-induced contribution. A rotation of the polar diagram away from the main crystallographic axis is also reported. **b** Temperature dependence of the integrated SHG intensity. **c** Temperature dependence of the calculated parameter κ, for two regions of interest.  **d** PL measurements of temperature-dependent spin-valley polarization for the neutral (black) and charged (red) exciton in the same sample.

**Fig. 5 Valley population imbalance imaging for monolayer and few-layer WS$_2$.**

**a** Optical image of a WS$_2$ flake consisting of monolayer, few-layer and bulk regions **b** SHG intensity mapping of the WS$_2$ flake at 78K, **c** Imaging of VPI in terms of the dimensionless parameter κ at 78 K. **d-f** Polar diagrams of SHG intensity with varying angle φ of incident field polarization at, 78, 100 and 200 K for monolayer and **g-i** few-layer regions. Red spheres correspond to experimental data and the blue line to the fitting with equation (4).

**Fig. 6 Comparison of valley population imbalance between atomically thin WS$_2$ and WSe$_2$.**

a Optical image of a large area comprising monolayer and few-layer WS$_2$ as well as monolayer WSe$_2$. The corresponding areas are encircled by the red dashed lines **b** SHG intensity imaging at 78 K **c** VPI imaging and **d** histogram in terms of the dimensionless parameter κ at 78 K.

**Figures**

**Figure 1**

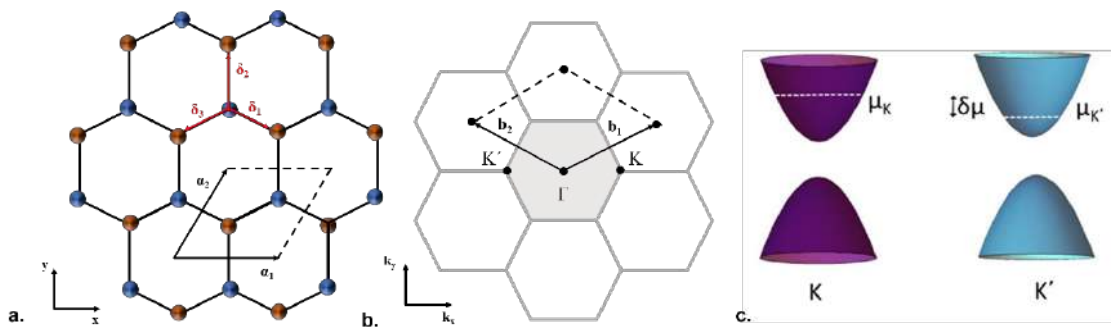

**Figure 2**

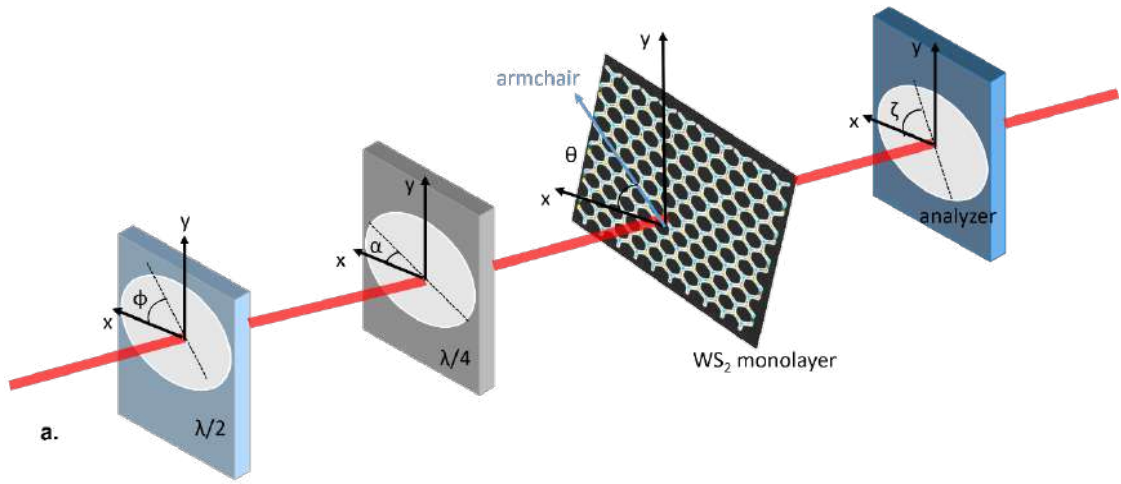

a.

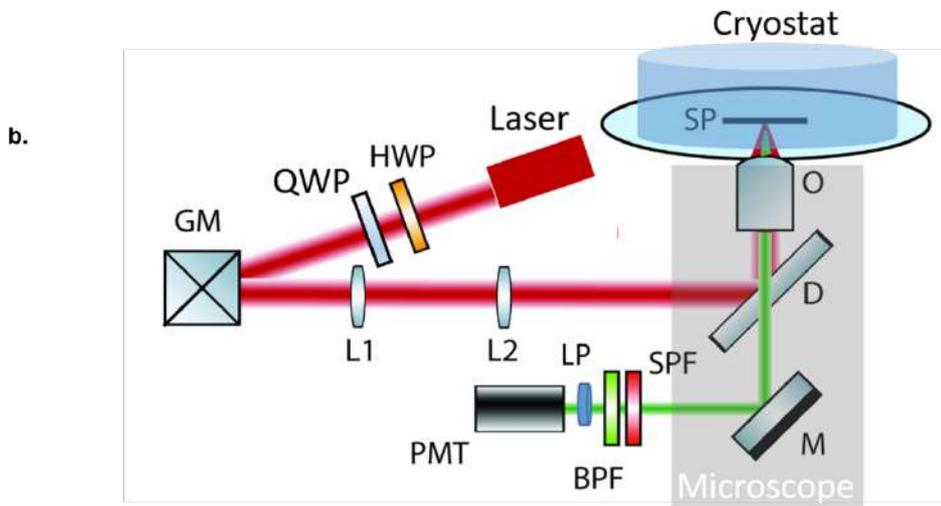

b.

**Figure 3**

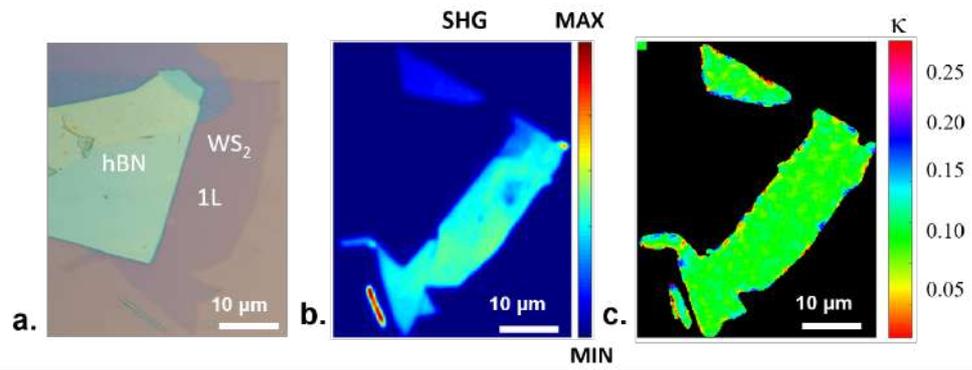

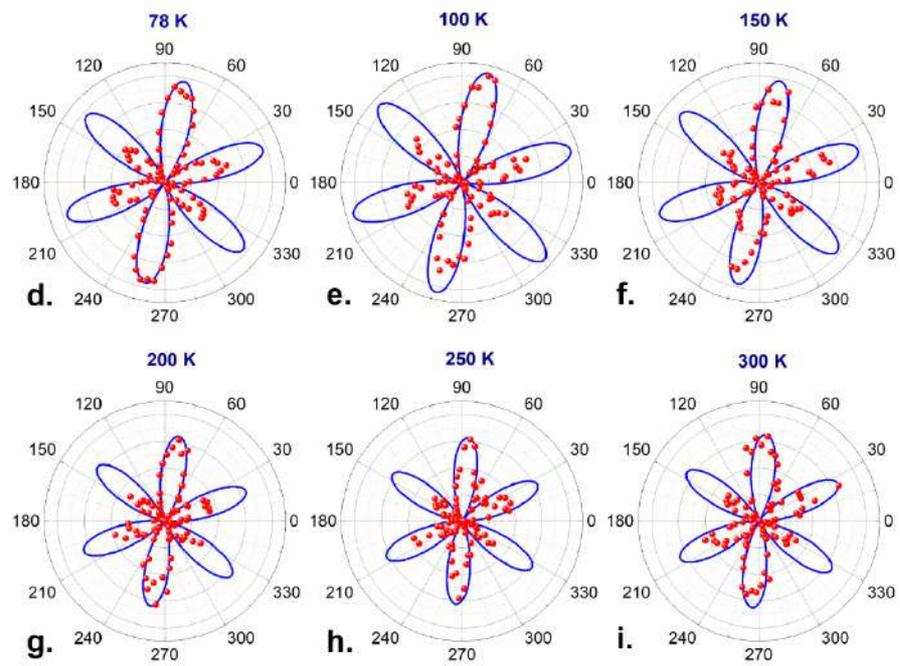

**Figure 4**

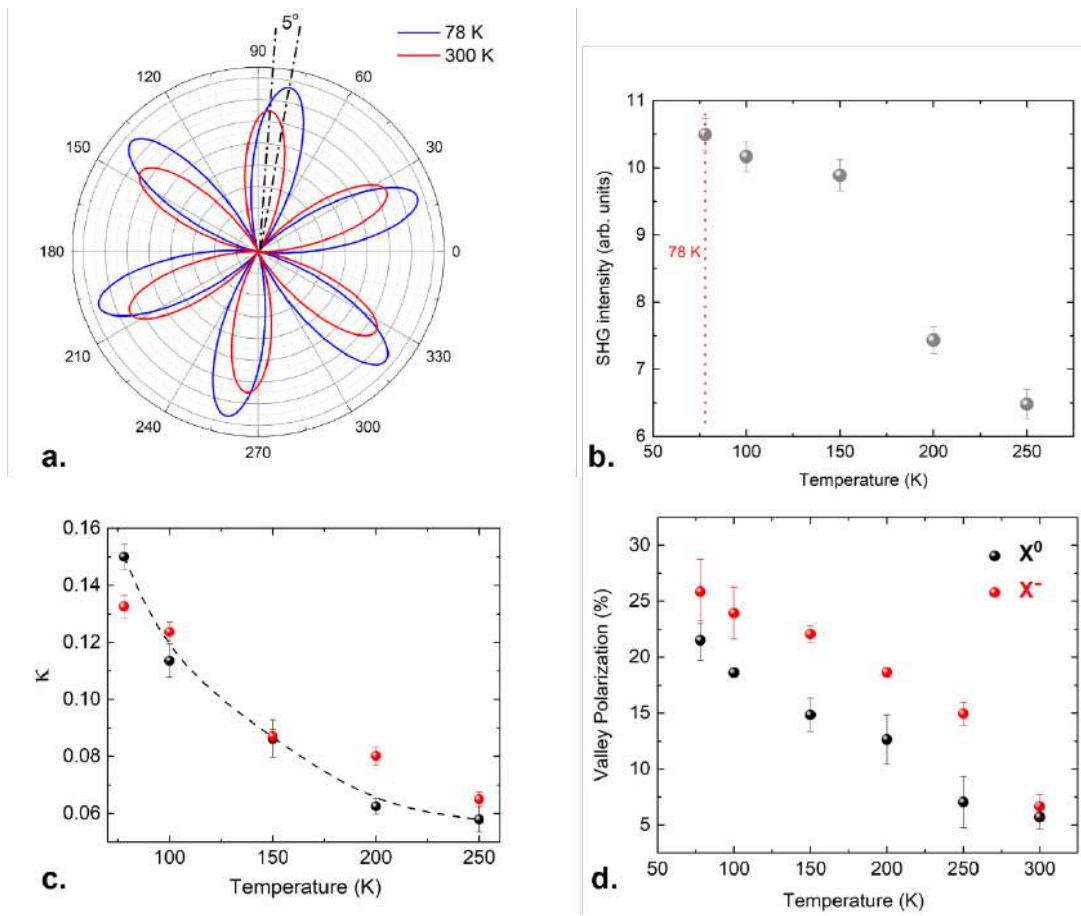

**Figure 5**

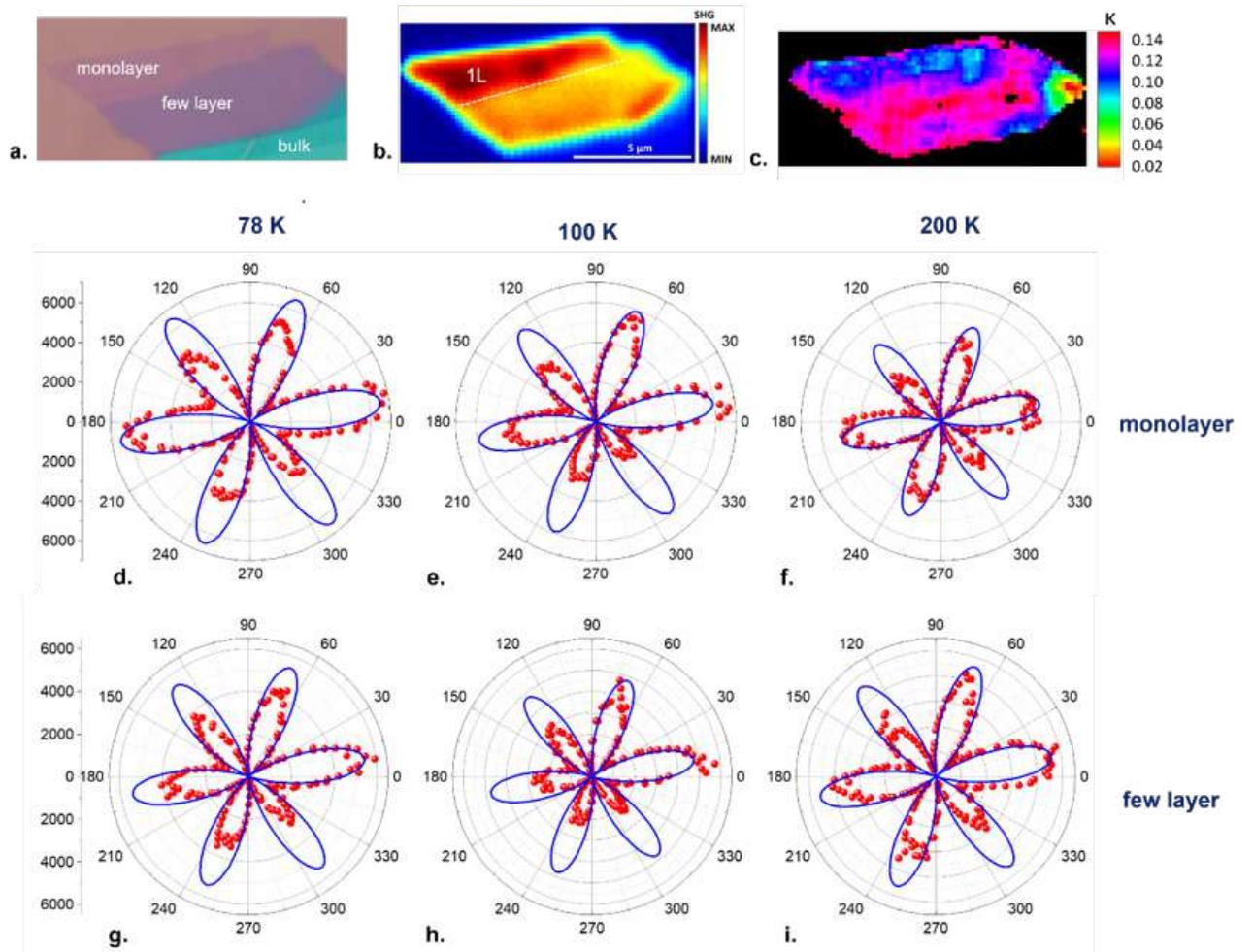

**Figure 6**

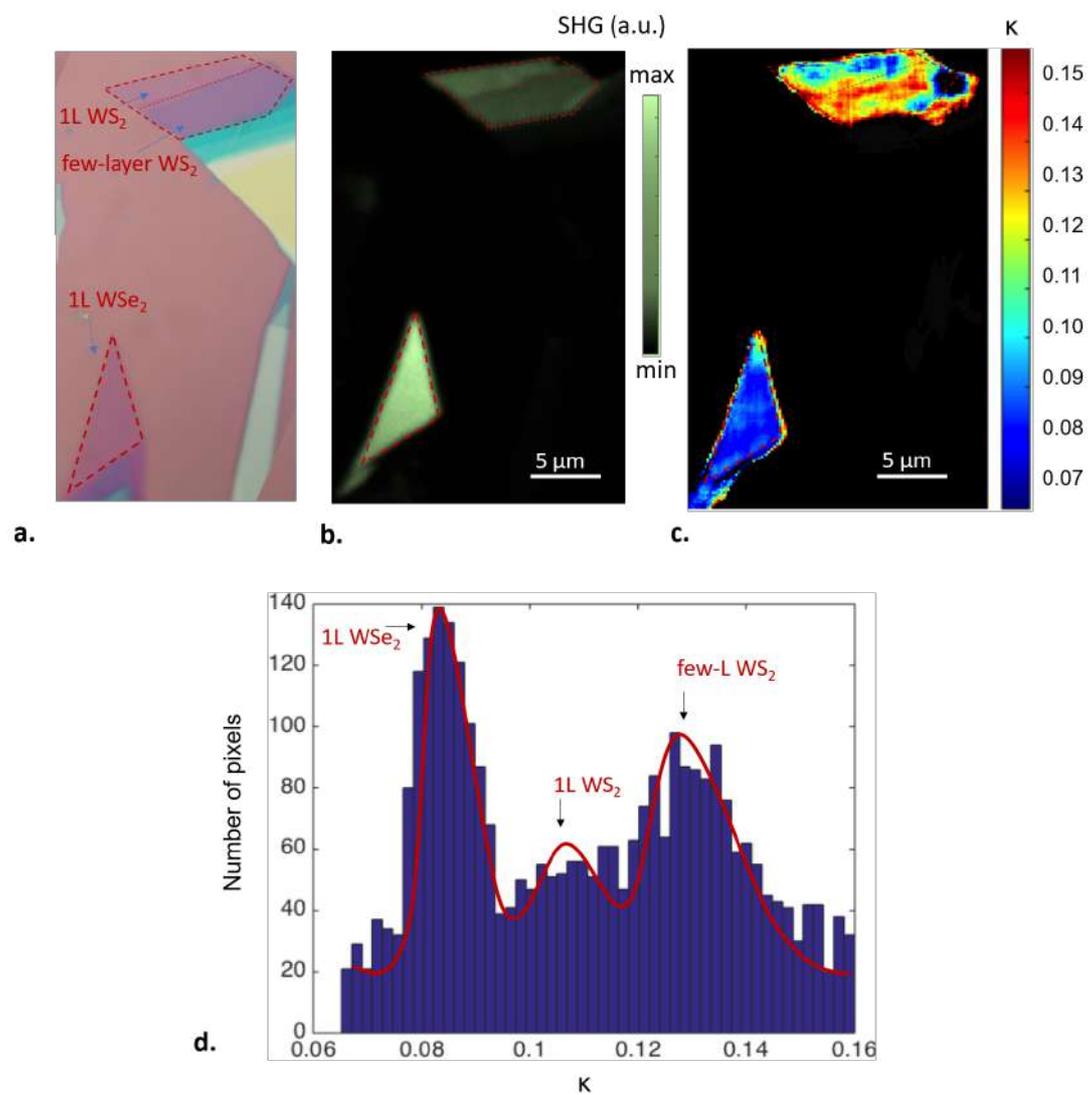